\documentstyle[epsfig]{mn}

\newif\ifAMStwofonts

\ifoldfss
  \ifCUPmtlplainloaded \else
    \NewTextAlphabet{textbfit} {cmbxti10} {}
    \NewTextAlphabet{textbfss} {cmssbx10} {}
    \NewMathAlphabet{mathbfit} {cmbxti10} {} 
    \NewMathAlphabet{mathbfss} {cmssbx10} {} 
  \fi
  \ifAMStwofonts
    \ifCUPmtlplainloaded \else
      \NewSymbolFont{upmath} {eurm10}
      \NewSymbolFont{AMSa} {msam10}
      \NewMathSymbol{\upi}     {0}{upmath}{19}
      \NewMathSymbol{\umu}     {0}{upmath}{16}
      \NewMathSymbol{\upartial}{0}{upmath}{40}
      \NewMathSymbol{\leqslant}{3}{AMSa}{36}
      \NewMathSymbol{\geqslant}{3}{AMSa}{3E}

       \let\ge=\geqslant
    \fi
  \fi
\fi 

\ifnfssone
  \newmathalphabet{\mathit}
  \addtoversion{normal}{\mathit}{cmr}{m}{it}
  \addtoversion{bold}{\mathit}{cmr}{bx}{it}
  \newmathalphabet{\mathbfit} 
  \addtoversion{normal}{\mathbfit}{cmr}{bx}{it}
  \addtoversion{bold}{\mathbfit}{cmr}{bx}{it}
  \newmathalphabet{\mathbfss} 
  \addtoversion{normal}{\mathbfss}{cmss}{bx}{n}
  \addtoversion{bold}{\mathbfss}{cmss}{bx}{n}
  \ifAMStwofonts
    \ifCUPmtlplainloaded \else
      \UseAMStwoboldmath
      \makeatletter
      \new@mathgroup\upmath@group
      \define@mathgroup\mv@normal\upmath@group{eur}{m}{n}
      \define@mathgroup\mv@bold\upmath@group{eur}{b}{n}
      \edef\UPM{\hexnumber\upmath@group}
      \new@mathgroup\amsa@group
      \define@mathgroup\mv@normal\amsa@group{msa}{m}{n}
      \define@mathgroup\mv@bold\amsa@group{msa}{m}{n}
      \edef\AMSa{\hexnumber\amsa@group}
      \makeatother
      \mathchardef\upi="0\UPM19
      \mathchardef\umu="0\UPM16
      \mathchardef\upartial="0\UPM40
      \mathchardef\leqslant="3\AMSa36
      \mathchardef\geqslant="3\AMSa3E

       \let\ge=\geqslant
    \fi
  \fi
\fi 

\ifnfsstwo
  \DeclareMathAlphabet{\mathbfit}{OT1}{cmr}{bx}{it}
  \SetMathAlphabet\mathbfit{bold}{OT1}{cmr}{bx}{it}
  \DeclareMathAlphabet{\mathbfss}{OT1}{cmss}{bx}{n}
  \SetMathAlphabet\mathbfss{bold}{OT1}{cmss}{bx}{n}
  \ifAMStwofonts
    \ifCUPmtlplainloaded \else
      \DeclareSymbolFont{UPM}{U}{eur}{m}{n}
      \SetSymbolFont{UPM}{bold}{U}{eur}{b}{n}
      \DeclareSymbolFont{AMSa}{U}{msa}{m}{n}
      \DeclareMathSymbol{\upi}{0}{UPM}{"19}
      \DeclareMathSymbol{\umu}{0}{UPM}{"16}
      \DeclareMathSymbol{\upartial}{0}{UPM}{"40}
      \DeclareMathSymbol{\leqslant}{3}{AMSa}{"36}
      \DeclareMathSymbol{\geqslant}{3}{AMSa}{"3E}

       \let\ge=\geqslant
    \fi
  \fi
\fi 

\ifCUPmtlplainloaded \else
  \ifAMStwofonts \else 
    \def\upi{\pi}
    \def\umu{\mu}
    \def\upartial{\partial}
  \fi
\fi

\title{The Ages and Metallicities of Early Type Galaxies in the Fornax Cluster}
\author[Harald Kuntschner \& Roger L. Davies]
       {Harald Kuntschner \& Roger L. Davies \\
        University of Durham, Durham DH1 3LE, England}
\date{accepted ... received ...}

\pagerange{\pageref{firstpage}--\pageref{lastpage}}
\pubyear{1997}

\begin{document}

\maketitle

\label{firstpage}

\begin{abstract}
  We have measured central line strengths for a complete sample of
  early type galaxies in the Fornax cluster, comprising 11 elliptical
  and 11 lenticular galaxies, more luminous than $M_B=-17$. In contrast
  to the elliptical galaxies in the sample studied by Gonz\'{a}lez (and
  recently revisited by Trager) we find that the Fornax ellipticals
  follow the locus of galaxies of fixed age in Worthey's models and
  have metallicities varying from roughly solar to three times solar.
  The lenticular galaxies however exhibit a substantial spread to
  younger luminosity weighted ages indicating a more extended star
  formation history. We present measurements of the more sensitive
  indices: C4668 and H$\gamma_A$; these confirm and re-enforce the
  conclusions that the elliptical galaxies are coeval and that only the
  lenticular galaxies show symptoms of late star-formation. The
  inferred difference in the age distribution between lenticular and
  elliptical galaxies is a robust conclusion as the models generate
  consistent relative ages using different age and metallicity
  indicators even though the absolute ages remain uncertain. The young
  luminosity weighted ages of the S0s in the Fornax cluster are
  consistent with the recent discovery that the fraction of S0 galaxies
  in intermediate redshift clusters is a factor of 2-3 lower than found
  locally and suggests that a fraction of the cluster spiral galaxy
  population has evolved to quiescence in the 5 Gyr interval from z=0.5
  to the present. Two of the faintest lenticular galaxies in our sample
  have blue continua and strong Balmer-line absorption suggesting
  starbursts $\la$2 Gyrs ago. These may be the low redshift analogues
  of the starburst or post-starburst galaxies seen in clusters at
  z=0.3, similar to the H$\delta$ strong galaxies in the Coma cluster.
\end{abstract}


\section{INTRODUCTION}
The conventional view that luminous elliptical galaxies are old, coeval
and created about 15 Gyrs ago has been established over many decades.
In this picture the global spectrophotometric relations observed for
ellipticals, for example the colour-magnitude relation
(Sandage~\&~Visvanathan~1977; Bower, Lucey \& Ellis 1992) are accounted
for by the steady increase in the abundance of heavy elements with
increasing galaxy mass which arises naturally in galactic wind models
such as that of Arimoto \& Yoshii (1987) and Kodama \& Arimoto (1997).
This view has received support from the small scatter observed in the
Fundamental Plane (Renzini~\&~Ciotti,~1993) and from the small scatter
in the Mg$_2$-$\sigma$ relation (Bender, Burstein \& Faber 1993) both
of which appear to be difficult to establish if there is any
significant age spread amongst elliptical galaxies. Recent observations
have however challenged this conventional interpretation of the data
(Gonz\'{a}lez 1993, hereafter G93; Faber~et~al.~1995) and suggested
that large age variations may be present amongst elliptical galaxies.
The integrated light spectral energy distributions derived from
single-age, single-metallicity models of early type galaxy spectra show
that the broad band colours and the widely used Mg$_2$ index are
largely degenerate in age and metallicity making these parameters
difficult to determine independently. Worthey (1994, hereafter W94)
however identified spectral features that are largely sensitive to age
and metallicity individually and was thus able to determine these
parameters.

These studies are based on measurements of the Faber-Burstein indices
defined initially for the Lick/IDS spectra and described in W94.  The
age sensitive absorption features are the Balmer lines, in particular
H$\beta$. The metallicity indicators are iron and magnesium absorption
features, in particular the [MgFe] index (defined in G93) which
combines two strong iron lines and the Mgb feature.  The 41 elliptical
galaxies studied by Gonz\'{a}lez have a large range in H$\beta$
absorption strength and a limited range in metal line strength [MgFe].
Combined with Worthey's models these indicate a large range in age,
from $\sim$2 to $\ge$12 Gyrs, with a modest spread in metallicity from
solar to roughly three times solar (see Fig.~\ref{fig:HbMgFe}b).

Jones \& Worthey (1995) identified more sensitive features for
metallicity and age: the C4668 feature and H$\gamma_{HR}$ measured at
high resolution. Trager (1997) recently revisited the G93 data and
analysed the original Lick sample using C4668 and the new higher order
Balmer line indices modelled by Worthey \& Ottaviani (1997, hereafter
WO97) to extend the application of the models with greater certainty.
He confirmed the G93 result and ascribed the differences in the
absolute ages of galaxies derived from different pairs of indices to
the well known over-abundance (compared to the solar ratios) of
magnesium compared to iron in luminous ellipticals (Peletier 1989;
Worthey, Faber \& Gonz\'{a}lez 1992; Davies, Sadler \& Peletier 1993;
Greggio 1997).

Gonz\'{a}lez's sample includes galaxies that are largely drawn from
relatively low density environments with a few galaxies taken from
nearby clusters. It was not intended to be a complete sample.  Here we
present a {\it complete} sample of early type galaxies in the Fornax
cluster brighter than $M_B=-17$.  In section~2 we describe the
observations and data analysis. In section~3 we present our
measurements of line strengths in the Fornax galaxies and make a direct
comparison of these with Gonz\'{a}lez sample.  We then apply the new,
more precise, age/metallicity indices and show that these re-enforce
our conclusions that the Fornax ellipticals are coeval and that ongoing
star-formation occurs in the S0 galaxies. We briefly discuss two
galaxies with remarkably blue spectra before bringing together our
conclusions in section~4 where we also speculate on the implications
for the role of morphology and environment in the star-formation
history of early type galaxies.

\section{OBSERVATIONS AND DATA REDUCTION} 
Our sample of 22 early type galaxies have been selected from the
catalogue of Fornax galaxies, Ferguson (1989, hereafter F89), in order
to obtain a complete sample down to $B_T = 14.2$ or
$M_B=-17$\footnote{Adopting a distance modulus of $m-M=31.2$}. We have
adopted the morphological classifications given by F89 and checked them
with images we obtained on the Siding Spring 40$\arcsec$ telescope.
From these we noted a central dust lane in ESO359-G02 and a central
disk in ESO358-G59 which led us to classify them as lenticular
galaxies. We classified IC2006 as elliptical, as it was not classified
by F89. NGC 1428 was not observed because of the bright star close to
its centre.  The observations were carried out with the AAT (3.9m) on
the nights of 1996 December 6-8 using the RGO spectrograph equipped
with a Tek 1K detector. We used the 600V grating resulting in a useful
wavelength range from 4243 \AA~to 5828 \AA. The slit width of 2.3
arcsec produced a spectral resolution of 4.1~\AA~(FWHM). One pixel
along the slit spanned 0.77 arcsec.  Typically, exposure times were
between 300 and 1800 sec per galaxy. The slit was centred on the
nucleus at $PA=90^\circ$. The seeing was generally better than one
arcsec.  Additionally we observed 15 different standard stars (mainly
K-giants) during twilight to act as templates for velocity dispersion
measurements as well as to calibrate our line-strength indices.
The flux standard GD~108 (Oke 1990) was observed to enable us to
correct the continuum shape of our spectra.

The standard data reduction procedures: flat-fielding, cosmic ray
removal, wavelength calibration, sky-subtraction and fluxing were
performed with IRAF. The central spectrum for each galaxy was extracted
by fitting a low order polynomial to the position of the centre along
the wavelength direction, re-sampling the data in the spatial direction
and finally co-adding the spectra within a 3.85 arcsec aperture
($=5$~pixel). The spectra, logarithmically rebinned in wavelength, were
used to derive redshifts and central velocity dispersions using a
simple Fourier Cross-correlation method ({\tt fxcor} in IRAF).

We measured line-strengths for [MgFe], C4668, H$\beta$ and H$\gamma_A$
in the Lick/IDS system described in detail in W94 \& WO97. The
pass-bands we used are shown overplotted on example spectra in
Fig.~\ref{fig:spec}.  The transformation from the observed system to
the Lick/IDS system was performed following previous authors and the
suggestions by WO97. In particular the fluxed spectra were artificially
broadened with a Gaussian of wavelength dependent width, such that the
Lick resolution was best matched at each wavelength (see Fig. 7 in
WO97). We corrected our indices for velocity dispersion using broadened
star spectra (see eg. Davies, Sadler \& Peletier 1993). Using stars and
galaxies we observed in common with the Lick/IDS data (Trager 1997) we
established small offsets to bring our measurements onto the Lick
system.

\section{RESULTS}

\subsection{The H$\beta$ {\it vs} [MgFe] diagram}
G93 successfully used a combination of H$\beta$ and [MgFe] to
disentangle the effects of age and metallicity. To make a direct
comparison we will first plot our data in the same co-ordinates as
Gonz\'{a}lez and explore the use of improved indices in section 3.2.

\begin{figure*}
    \leavevmode
      \epsfig{file=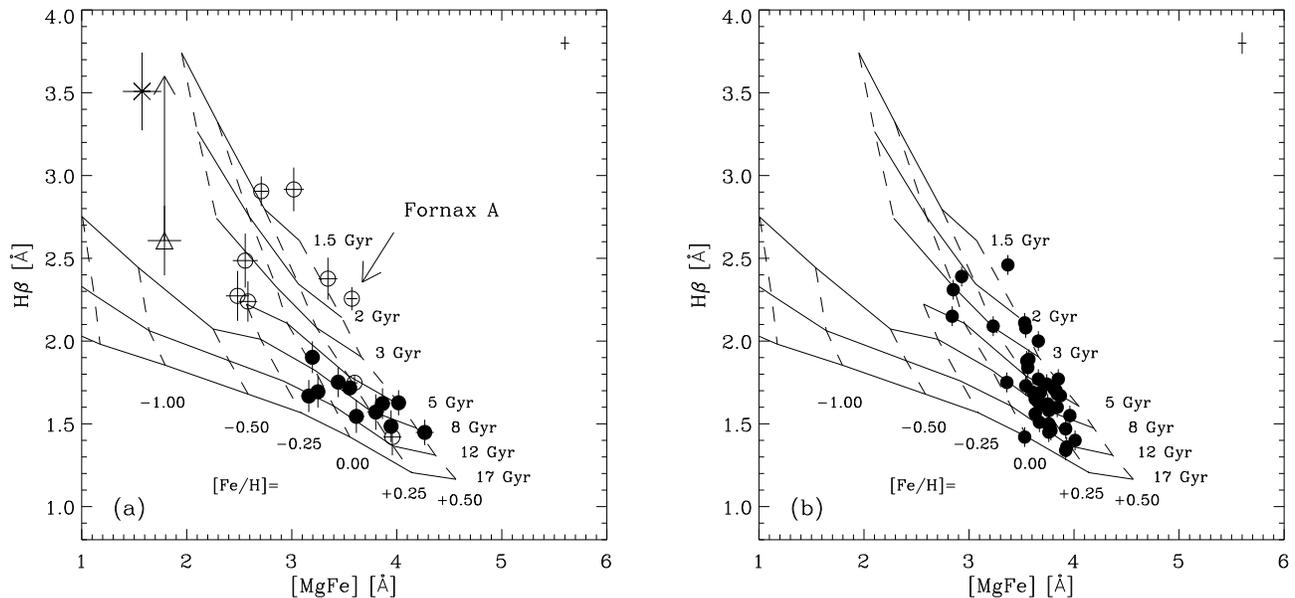,height=8.5cm, width=18cm}    
    \caption{(a) H$\beta$ equivalent width {\it vs} [MgFe] equivalent width
      diagram for the complete sample of Fornax early type galaxies.
      Overplotted are models by W94. Filled circles and open circles
      represent ellipticals and S0s respectively. The cross and open
      triangle represent possible post-starburst and starburst galaxies
      respectively. The cross in the right upper corner indicates the
      rms uncertainty in the transformation to the Lick/IDS system.
      The arrow attached to the galaxy ESO358-G25 (open triangle)
      indicates an emission correction, for details see text. (b) The
      Gonz\'{a}lez (1993) sample (R$_e/8$ aperture) in the same
      coordinates as in (a).  Note that these elliptical galaxies show
      a large range in age and a modest range in metallicity.}
    \label{fig:HbMgFe}
\end{figure*}

In Fig.~\ref{fig:HbMgFe}a we present a plot of H$\beta$ equivalent
width {\it vs} [MgFe] equivalent width for our 22 galaxies. The error
bars on the data points represent the photon noise and the error in the
velocity dispersion correction. The elliptical galaxies in Fornax
(filled circles) have weak H$\beta$ absorption spanning a modest range
in [MgFe]. The S0s (open circles) on the other hand span a larger range
of values in this diagram, typically having stronger H$\beta$
absorption.

Following G93 we overplot predictions from single-burst stellar
population models (W94). Although the absolute age calibration may be
insecure, we see that our sample of Fornax ellipticals are old and of
similar age. According to the models the metallicity ranges from just
sub-solar to about three times solar. The Fornax S0 galaxies however
have much lower luminosity weighted ages and a greater range in
metallicity.  This is very much in contrast to what G93 found. His
sample is shown in Fig. \ref{fig:HbMgFe}b; the galaxies exhibit a large
spread in luminosity weighted age from greater than 12~Gyrs to less
than 2~Gyrs, in fact Gonz\'{a}lez' sample looks much more like the
Fornax S0s than the Fornax ellipticals. Only four galaxies of the G93
sample are classified as disk galaxies in the RC3 (de Vaucouleurs et
al. 1991).

We interpret the stronger H$\beta$ absorption found in S0s as
indicating a younger `mean' age, but recall that line-strength indices
reflect only the integrated, luminosity weighted, properties in a
galaxy. As young populations tend to be much more luminous than old
ones, a small (in mass) young population can dramatically change the
strength of indices, in particular H$\beta$. de Jong \& Davies (1997)
and Trager (1997) have demonstrated this and showed that the high
H$\beta$ galaxies in G93 tend to have disky isophotes.  de Jong \&
Davies suggested that the ongoing star-formation might be associated
with the presence of a disk.  In our sample, Fornax A, often cited as
the product of a recent merger, might be a typical example of an
``old'' galaxy which looks young due to ongoing star-formation in a
disk.  Less dramatically the same effect could affect many lenticular
galaxies.

\subsection{\label{sec:hga}New Indices : C4668 \& H$\gamma_A$}

Here we discuss the application of a more sensitive and more accurately
determined metallicity index C4668, and an age index, H$\gamma_A$, that
is also more precisely determined and less sensitive to contamination
by emission.

Mg appears to be overabundant compared to Fe in luminous elliptical
galaxies whereas the models are based on solar abundance ratios. As a
result of this the use of different metallicity indices can result in
quite different estimates for the absolute ages of the same galaxies.
Worthey (1995) identified C4668 as a particularly sensitive metallicity
feature that, while overabundant compared to Fe, is less overabundant
than Mg. He points out however that using a different metallicity
indicator does not change the relative distribution of the ages and
metallicities of galaxies much but simply shifts the distribution of
all galaxies together so that their relative ages should be insensitive
to the choice of diagnostic.
 
G93 found nebular emission in more than half of the galaxies in his
sample. The strength of the {\it stellar} absorption at H$\beta$ is
therefore uncertain and requires a correction for the estimated
infilling due to emission. Gonz\'{a}lez adopted an empirical
prescription based on the strength of the OIII emission but the
correctness of this has been challenged by Carrasco (1996) who proposes
that no correction should be made. We did not attempt to correct
H$\beta$ for emission, rather we used H$\gamma_A$ which is less
sensitive to contamination by emission. The relative strength of
nebular emission decreases rapidly with the order of the Balmer-line
(Osterbrock 1989) so that the dilution effect is much reduced.
H$\gamma_A$ is a more sensitive age indicator than H$\beta$ because the
models predict a much wider range in equivalent width for the same age
difference. In addition H$\gamma_A$ is more precisely determined as (i)
the wide sidebands produce improved photon statistics, (ii) there is a
smaller rms error in the dispersion velocity corrections.

\begin{figure}
    \leavevmode
      \epsfig{file=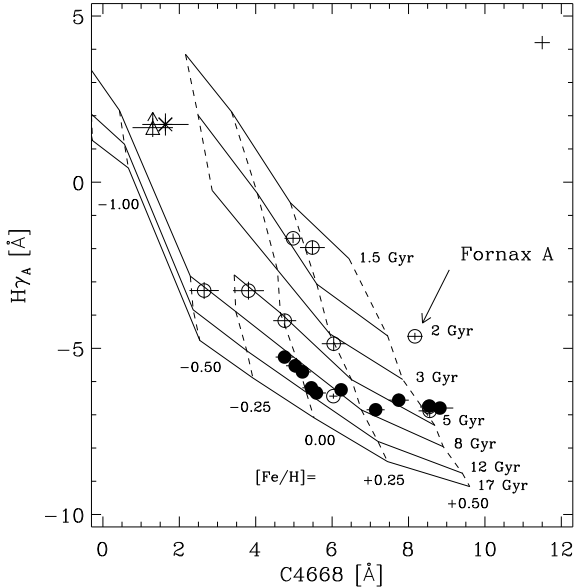,height=8.5cm, width=8.5cm}    
    \caption{H$\gamma_A$ equivalent width {\it vs} C4668 equivalent width
      diagram for a complete sample of Fornax early type galaxies.
      Symbol definitions as in Fig.~\ref{fig:HbMgFe}a. Overplotted are
      models by WO97. The negative values in H$\gamma_A$ {\em do not}
      indicate emission but are created entirely by the definition of
      the pseudo-continuum bands (see WO97). Note that ESO358-G25 (open
      triangle) moves to a much younger age with respect to
      Fig.~\ref{fig:HbMgFe}a. The ellipticals with metallicities
      greater than solar have roughly constant H$\gamma_A$.}
    \label{fig:HgaC4668}
\end{figure}

In Fig.~\ref{fig:HgaC4668} we present a plot of H$\gamma_A$ equivalent
width {\it vs} C4668 equivalent width. The symbol definitions are the
same as in Fig.~\ref{fig:HbMgFe}a. The subscript ``A'' on H$\gamma$
indicates a ``wide'' ($\sim$40~\AA) central passband. We have
overplotted new model predictions (WO97). In this diagram the
ellipticals follow a tight relation at low H$\gamma_A$ values varying
mostly in C4668. S0s are distributed to much higher values of
H$\gamma_A$ absorption. The distribution is similar to that in the
H$\beta$ {\it vs} [MgFe] diagram and the precision of both age and
metallicity is increased. Remarkably the scatter of elliptical galaxies
is reduced and for metallicities greater than solar they exhibit
more-or-less constant H$\gamma_A$. The new, more accurate, indicators
confirm our principal result that the lenticular galaxies in the Fornax
cluster have lower luminosity weighted ages than the elliptical
galaxies which appear to be roughly coeval and vary mainly in
metallicity.

There is a hint that the most metal rich ellipticals seem to be younger
by $\sim$3 Gyrs than their metal poor brethren.  While the absolute age
calibration is not secure the relative ages should be valid,
nevertheless the apparent reduction in the luminosity weighted ages of
the most luminous galaxies may be an artifact in the models.

\subsection{Two post-starburst or starburst galaxies}

\begin{figure*}
    \leavevmode
      \epsfig{file=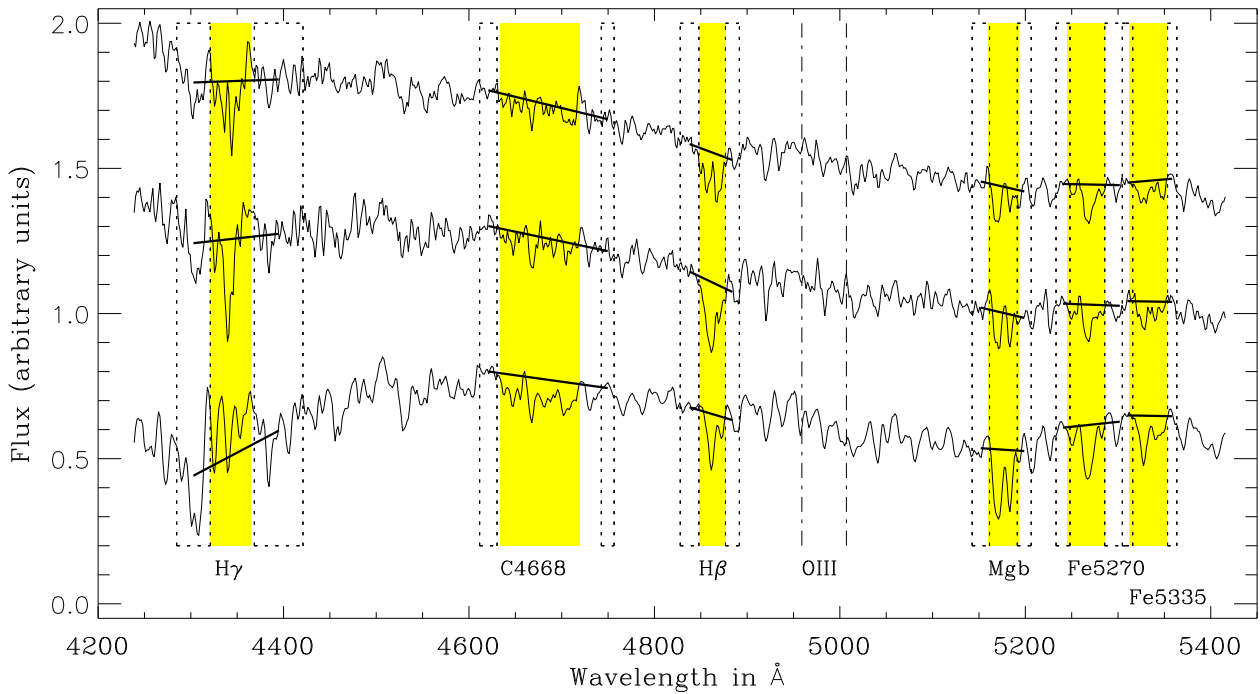,height=10cm, width=18cm}    
    \caption{The spectra of ESO358-G25, ESO359-G02 and NGC1336 (from
      the top) are shown. The shaded areas mark the central passbands
      of the indices considered in this letter. The thick solid lines
      indicate the pseudo-continuum defined by two side-passbands
      (dashed boxes). We have also marked the position where OIII
      emission at 4959~\AA~\& 5007~\AA~would be found if present. The
      spectra are {\em not} broadened to the Lick resolution in order
      to illustrate the different line strengths.  Note the emission in
      H$\beta$ and H$\gamma$ for ESO358-G25 and that the spectra of
      both ESO-galaxies do not drop bluewards of 4500~\AA~but stay
      constant.}
    \label{fig:spec}
\end{figure*}

One of the most striking differences in using H$\gamma_A$ instead of
H$\beta$ is seen in the behaviour of ESO358-G25 represented by an open
triangle. This galaxy moves to much younger ages compared to the rest
of the sample. In fact it shows emission in H$\beta$ and H$\gamma$
filling in the absorption (see Fig.~\ref{fig:spec}). The arrows in
Figs.~\ref{fig:HbMgFe}a \& \ref{fig:HgaC4668} indicate an estimated
emission-correction determined by a rough subtraction of the emission
features. ESO358-G25 is the only galaxy in our sample which shows
obvious Balmer emission and as a result it is the only galaxy which
moves to significantly younger ages or greater relative index values in
Fig.~\ref{fig:HgaC4668}. The two lenticular galaxies ESO358-G25 (open
triangle) and ESO359-G02 (cross) have remarkable spectra for early type
galaxies, they show blue continua, strong Balmer lines, weak metal
lines and there is a hint of weak OIII emission (see
Fig.~\ref{fig:spec}). These galaxies are amongst the faintest in our
sample and are $\sim$3 degrees away from the centre of the cluster.
They are reminiscent of the high H$\delta$ galaxies in the Coma cluster
found by Caldwell et al. (1993) or the galaxies in redshift z=0.3
clusters identified by Couch and Sharples (1987) and by Barger et al.
(1996) as being in the post-starburst phase (ESO359-G02) or starburst
phase (ESO358-G25).

\section{CONCLUSIONS and SPECULATIONS}

\subsection{Conclusions}

We have measured the central line strength indices in a complete sample
of early type galaxies brighter than $M_B=-17$ in the Fornax cluster
and have applied the models of W94 and WO97 to estimate their ages and
metallicities. We find that:

\begin{enumerate}
\item The elliptical galaxies appear to be roughly coeval although
  their absolute ages remain uncertain. They form a sequence in
  metallicity varying from just below solar metallicity to about three
  times solar metallicity.  This result is consistent with the
  conventional view of old, coeval elliptical galaxies and contrasts
  remarkably with the results from the G93 sample.
  
\item The lenticular galaxies have metallicities ranging from one-third
  to three times solar metallicity and {\it have luminosity weighted
    ages that are much younger than those of the ellipticals}, spanning
  from less than $\sim$2 Gyrs to 12 Gyrs. In this respect the
  distribution of G93 galaxies follows that of Fornax S0s more closely
  than the Fornax ellipticals.
  
\item We have discovered that two of the fainter lenticular galaxies
  appear to have undergone star-formation in the last 2 Gyrs (in one
  case very much more recently). These appear to be the low redshift
  analogues to the post-starburst or starburst galaxies seen by Couch
  \& Sharples (1987) and recently by Barger et al. (1996) in redshift
  z=0.3 clusters. We note that, like Fornax A, these galaxies lie on
  the periphery of the cluster.

\end{enumerate}

\subsection{Speculations}
We are now in a position to speculate on the role of galaxy morphology
and environment in the star-forming history of early-type galaxies.

\begin{enumerate}
  
\item In the Fornax cluster only lenticular galaxies exhibit symptoms
  of recent star-formation. Is it possible that this is a general
  result? This would be consistent with the findings of Dressler et al.
  (1997), that in clusters at z=0.5 the fraction of elliptical galaxies
  remains constant but the fraction of S0 galaxies is 2-3 times lower
  than is found in local clusters. They suggest that a fraction of the
  spiral galaxy population has evolved to quiescence in the 5 Gyr
  interval from z=0.5 to the present. This would indeed produce a
  cluster population of youthful S0 galaxies as we observe. The
  occurence of late star-formation only in S0 galaxies supports the
  view that early type galaxies appear young because a modest mass
  fraction of young stars influence their spectra, rather than that
  they formed recently.~de Jong and Davies (1997) and Trager (1997)
  have noted the tendency for high H$\beta$, or young galaxies to have
  disky isophotes re-enforcing the suggestion that they may originate
  as exhausted spirals. It seems possible that at least {\it some} of
  the young galaxies in the G93 sample may be mis-classified lenticular
  galaxies.  We are undertaking an imaging survey of the G93 sample to
  explore this.
  
\item Alternatively it may be that elliptical galaxies in dense
  environments are older and show less age spread than those in loose
  groups and the field. This has certainly been suggested by others eg.
  Larson, Tinsley \& Caldwell (1980) and Guzman \& Lucey (1993). If we
  interpret the difference in the age distribution between the G93 and
  Fornax samples as reflecting the difference in environment between
  field and cluster galaxies we should also recall that the fraction of
  early type galaxies that have disks increases dramatically at lower
  densities (Dressler 1980). There is an urgent need to construct
  complete datasets for other nearby clusters and to construct a well
  defined low density sample to test whether it is morphology,
  environment or both that generate the age differences that we see.
  
\item We note that in the H$\gamma$ {\it vs} C4668 diagram the most
  metal-rich (luminous) galaxies appear to be younger by $\sim$3 Gyrs
  than the galaxies of lower metallicity. It is possible that this age
  difference is an artifact of the models. However if true this
  supports the hierarchical picture for the construction of galaxies in
  which massive galaxies are the last to be assembled and are therefore
  the youngest.
\end{enumerate}

\section{ACKNOWLEDGEMENTS}

We thank PATT for the allocation of telescope time and the staff of the
AAO for supporting our observing. We also thank Matthew Colless, Ken
Freeman and the staff of the Siding Spring Observatory for the use of
the 40$\arcsec$ telescope and for their assistance. We acknowledge
interesting discussions of this work with Ian Smail, Roberto Saglia and
Doerte Mehlert. RLD wishes to thank the hospitality of the Lorenz
Centre and Prof. P.T. de Zeeuw during the 1997 Galaxies workshop. HK
wishes to thank the Dr.~Carl~Duisberg~Stiftung and University of Durham
for generous financial support.

\bsp

\label{lastpage}


\begin{thebibliography}{}

  \bibitem {} Arimoto N., Yoshii Y., 1987, AA, 173, 23
  
  \bibitem {} Barger A.J., Arag\'{o}n-Salamanca A., Ellis R.S., Couch
    W.J., Smail~I., Sharples R.M., 1996, MNRAS, 279, 1

  \bibitem {} Bender R., Burstein D., Faber S.M., 1993, ApJ, 411,153 
    
  \bibitem {} Bower R.G., Lucey J.R., Ellis R.S., 1992, MNRAS, 254, 601

  \bibitem {} Caldwell N., Rose J.A., Sharples R.M., Ellis R.S., Bower
    R.G., 1993, AJ, 106, 473 
    
  \bibitem {} Carrasco L., Buzzoni A., Salsa M., Recillas-Cruz E.,
    1996, in Buzzoni A., Renzini A., Serrano A., ed., ASP Conf. Ser.
    Vol. 86, Fresh Views of Elliptical Galaxies. Astron. Soc. Pac., San
    Francisco, p. 235
   
  \bibitem {} Couch W.J., Sharples R.M., 1987, MNRAS, 229, 423

  \bibitem {} Davies R.L., Sadler E.M., Peletier R.F., 1993, MNRAS,
    262, 650

  \bibitem {} de Jong R.S., Davies R.L., 1997, MNRAS, 285, 1L
    
  \bibitem {} de Vaucouleurs G., de Vaucouleurs A., Corwin, H.G., Jr.,
    Buta R.J., Paturel G., Fouqu\'{e}, 1991, Third Reference Catalogue
    of Bright Galaxies, (Springer Verlag), (RC3)
    
  \bibitem {} Dressler A., 1980, ApJ, 236, 351
    
  \bibitem {} Dressler A., Oemler A., Jr., Couch W.J., Smail I., Ellis
    R.S., Barger A., Butcher H., Poggianti B.M., Sharples R.M., 1997,
    preprint astro-ph/9707232

  \bibitem {} Faber S.M., Trager S.C., Gonz\'{a}lez J.J., Worthey G.,
    1995, in Van Der Kruit P.C., Gilmore G., ed., Proc. IAU Symp. 164,
    Stellar Populations. Kluwer, Dordrecht, p. 249

  \bibitem {} Ferguson H.C., 1989, AJ, 98, 367, (F89)    
    
  \bibitem {} Gonz\'{a}lez J.J., 1993, Ph.D. thesis, Univ.
    California, Santa Cruz, (G93)

  \bibitem {} Greggio L., 1997, MNRAS, 285, 151

  \bibitem {} Guzman R., Lucey J.R., 1993, MNRAS, 263, 47L 

  \bibitem {} Jones L.A., Worthey G., 1995, ApJL, 446, 31
    
  \bibitem {} Kodama T., Arimoto N., 1997, AA, 320, 41
    
  \bibitem {} Larson R.B., Tinsley B.M., Caldwell C.N., 1980, ApJ, 237,
    692

  \bibitem {} Osterbrock D.E., 1989, Astrophysics of Gaseous Nebulae
    and Active Galactic Nuclei, (Mill Valley, CA: University Science
    Books)

  \bibitem {} Oke J.B., 1990, AJ, 99, 1621
   
  \bibitem {} Peletier R.F., 1989, Ph.D. thesis, Univ. of Groningen,
    Groningen

  \bibitem {} Renzini A., Ciotti L., 1993, ApJL, 416, 49

  \bibitem {} Sandage A., Visvanathan, 1977, ApJ, 216, 214
    
  \bibitem {} Trager S.C., 1997, Ph.D. thesis, Univ. California, Santa
    Cruz

  \bibitem {} Worthey G., 1994, ApJS, 95, 107, (W94)
    
  \bibitem {} Worthey G., 1995, in Leitherer C., Fritze-von Alvensleben
    U., Huchra J., ed., ASP Conf. Series, Vol 98, From Stars to
    Galaxies: The impact of Stellar Physics on Galaxy Evolution.
    Astron. Soc. Pac., San Francisco, p. 467
    
  \bibitem {} Worthey G, Ottaviani D.L., 1997, ApJS, 111, 377, (WO97)

  \bibitem {} Worthey G., Faber S.M., Gonz\'{a}lez J.J., 1992, ApJ, 398,69 

\end{thebibliography}
\end{document}